\documentclass[aip,numerical,pop,preprint,amsmath,amssymb,letterpaper]{revtex4-1}

\usepackage{graphicx}
\usepackage{dcolumn}
\usepackage{bm}
\usepackage{color}
\usepackage{hyperref}
\hypersetup{colorlinks=true,citecolor=blue,linkcolor=black,urlcolor=blue}

\newcommand{\vek}[1]{\boldsymbol{#1}}
\newcommand{\uvek}[1]{\hat{\boldsymbol{#1}}}

\newcommand{\deltab}{\delta \vek{B}}
\newcommand{\bnorm}{\uvek{b}}

\newcommand{\mytensor}[1]{\underline{\boldsymbol{#1}}}

\newcommand{\mycite}[1]{[\onlinecite{#1}]}
\newcommand{\exb}{\boldsymbol{E}\times\boldsymbol{B}}

\begin{document}


\title{Gyrokinetic understanding of the edge pedestal transport driven by resonant magnetic perturbations in a realistic divertor geometry}
\author{R. Hager}
 \email{rhager@pppl.gov}
\affiliation{%
Princeton Plasma Physics Laboratory\\
P.O. Box 451, Princeton, NJ 08543, USA
}%
\author{C.S. Chang}%
\affiliation{%
Princeton Plasma Physics Laboratory\\
P.O. Box 451, Princeton, NJ 08543, USA
}
\author{N. M. Ferraro}%
\affiliation{%
Princeton Plasma Physics Laboratory\\
P.O. Box 451, Princeton, NJ 08543, USA
}
\author{R. Nazikian}%
\affiliation{%
Princeton Plasma Physics Laboratory\\
P.O. Box 451, Princeton, NJ 08543, USA
}

\preprint{The following article has been submitted to Physics of Plasmas.}
\preprint{After it is published, it will be found at \href{https://pop.aip.org}{https://pop.aip.org}.}

\date{\today}


\begin{abstract}
Self-consistent simulations of neoclassical and electrostatic turbulent transport in a DIII-D H-mode edge plasma under resonant magnetic perturbations (RMPs) have been performed using the global total-f gyrokinetic particle-in-cell code XGC, in order to study density-pump out and electron heat confinement.
The RMP field is imported from the extended magneto-hydrodynamics (MHD) code M3D-C1, taking into account the linear two-fluid plasma response.
With both neoclassical and turbulence physics considered together, the XGC simulation reproduces two key features of experimentally observed edge transport under RMPs: increased radial particle transport in the pedestal region that is sufficient to account for the experimental pump-out rate, and suppression of the electron heat flux in the steepest part of the edge pedestal.
In the simulation, the density fluctuation amplitude of modes moving in the electron diamagnetic direction increases due to interaction with RMPs in the pedestal shoulder and outward, while the electron temperature fluctuation amplitude decreases.
\end{abstract}


\maketitle

%
\section{Introduction}\label{sec:intro}

Application of resonant magnetic perturbations (RMPs) -- externally applied small amplitude 3D perturbations -- in tokamak magnetic confinement fusion devices is a widely used method for suppressing edge localized modes (ELMs) \cite{evans_2015} in high-confinement mode (H-mode) operation.
The capability to suppress ELMs may be essential for ITER, in which ELMs in high-current discharges can be projected to be so violent that they can seriously damage plasma-facing components \cite{loarte_2014}.
ITER relies on RMP coils for ELM control \cite{martin_2013}.
However, fundamental understanding at the kinetic level of RMP ELM control is still missing.

Evans \textit{et al.} provided an overview of the current state of RMP research in Ref. \mycite{evans_2015}.
Among the open questions are how RMPs penetrate the plasma \cite{hu_2019,fitzpatrick_2018,heyn_2008,ferraro_2012,beidler_2018,nardon_2010,park_2010}, and how RMPs and plasma interact to produce enhanced particle transport (density pump-out) without enhancing electron thermal transport \cite{taimourzadeh_2019,holod_2016,kwon_2018,hager_2019_1}. 
Density pump-out often reduces the central plasma density significantly (up to $50$\%) in experiments with RMP field \cite{evans_2006}.
Such a large pump-out level can significantly degrade fusion performance in ITER and increase the requirement for plasma fueling that can create other issues\cite{cui_2017}.
A kinetic level understanding of the density pump-out phenomenon is essential in order to reliably predict and improve the effect of RMPs on fusion performance in a reactor such as ITER.


Density pump-out by RMPs can occur with or without ELM suppression\cite{joseph_2012,nazikian_2015}, which implies that the phenonenon may be related to the physics inherent to RMPs rather than ELMs.
Also, the existence of measurable, large-size magnetic islands at the pedestal top does not appear to be a prerequesite for density pump-out, because pump-out often precedes the penetration of those islands \cite{nazikian_2015,hu_2019}.
Complicating the problem is the experimentally observed preservation of the edge electron heat barrier, which cannot be explained by the popular Rechester-Rosenbluth theory 
\cite{rechester_1978}
 that was developed for transport in fully stochastic magnetic fields.
This suggests that the RMP fields may not be completely stochastic in the tokamak edge.
In fact, previous papers found out that it is difficult to achieve a Rechester-Rosenbluth relevant, completely stochastic magnetic field in a tokamak plasma \cite{spizzo_2018,park_2010}.

A multiscale kinetic treatment may ultimately be required to address the complex multi-physics phenomena associated with an RMP model, including neoclassical and turbulent transport together.
Here we address these kinetic effects using total-f XGC gyrokinetic simulation \cite{ku_2018}. 

In this work, we revisit the same low-collisionality DIII-D H-mode discharge (\#157308) that was investigated in our previous, neoclassical-only study \cite{hager_2019_1}.
While that study found significantly increased particle transport in the thin
 weakly 
stochastic field layer at the pedestal foot around the separatrix, neoclassical transport alone proved insufficient to explain the density pump-out in the rest of the pedestal region.
The electron heat flux also increased in the thin stochastic layer, but only modestly, to a level,
 well below the fully stochastic Rechester-Rosenbluth level.
However, because neoclassical particle or electron heat transport is usually much smaller than turbulent transport, meaning that turbulence must be included for a more realistic investigation, no final conclusion could be reached in Ref. \onlinecite{hager_2019_1} with regard to the level of density pump-out or the persistence of the electron thermal transport barrier in the edge pedestal.
The present study self-consistently adds electrostatic turbulence physics to the previous neoclassical study as the first step.
The effects of the electromagnetic turbulence will be left for future work.

We use the total-f gyrokinetic particle-in-cell (PIC) code XGC \cite{ku_2018} including the spatially three-dimensional (3D) neoclassical and electrostatic turbulence physics together.
Our simulation takes into account the core heat and torque sources, the neutral particle recycling on the material wall with ionization and charge exchange reactions \cite{stotler_2017}, and a fully nonlinear Fokker-Planck-Landau collision operator suitable for the non-Maxwellian plasma in the narrow H-mode pedestal and scrape-off layer.
The RMP field used in the XGC simulation includes the linear plasma response and has been calculated with the extended MHD code M3D-C1.
We note here that a linear RMP penetration calculation from an MHD code can break down at a practical RMP level due to overlapping of the perturbed $\deltab$ from adjacent resonant surfaces \citenum{turnbull_2013,reiman_2015}
However, the same papers have also found that the linear solutions are qualitatively similar to the nonlinear solutions (see Fig. 7 of Ref. \onlinecite{reiman_2015}) at the practical RMP level even with the overlaping perturbations.
Thus, it is assumed here that the linear M3D-C1 calculation has the qualitatively correct magnetic perturbations.

The key outcomes regarding RMP-driven transport that we describe in this article are i) combined neoclassical and turbulent transport is necessary to explain the experimental-level density pump-out rate in the entire pedestal slope region;
and ii) the radial electron energy flux is dominated by the convective flux (i.e. flowing with the particle flux), and the turbulent electron heat flux is suppressed in the steepest part of the edge temperature pedestal inhibiting electron temperature relaxation.

The remainder of this article is organized as follows.
In Sec. \ref{subsec:model}, we discuss the numerical models used in XGC to compute the neoclassical and turbulent transport in the presence of RMPs.
The simulation setup is explained in Sec. \ref{subsec:case_setup}.
The results of our numerical study are analyzed in Sec. \ref{sec:results}.
In the first part of the analysis, Sec. \ref{subsec:turbulence}, we study the general properties such as fluctuation amplitudes and propagation direction of the turbulent modes observed in the simulations.
In the second part, \ref{subsec:transport}, we analyze the radial transport fluxes and how they change due to interaction with RMP field.
Summary and conclusion are provided in Sec. \ref{sec:summary}.

%
\section{Numerical model and test case setup}\label{sec:model}
\subsection{Numerical model}\label{subsec:model}
The numerical approach used in the present study is largely the same combination of extended MHD data and gyrokinetic simulation as in Ref. \mycite{hager_2019_1}.
Therefore, we only briefly review the key elements of this work-flow here and highlight the differences.
Note we use SI units throughout the remainder of this article, except for temperature representation in the figures.
First, the equilibrium magnetic field, including the three-dimensional RMP field from the DIII-D I-coils (two rows of six RMP coils on the high-field side, one above and one below the midplane and close to the inner wall) \cite{fenstermacher_2008}, is computed with M3D-C1 \cite{ferraro_2012,ferraro_2013,ferraro_2016}.

M3D-C1 first computes the axisymmetric equilibrium field by solving the Grad-Shafranov equation, and the vacuum RMP field generated by the I-coils by solving the Biot-Savart law.
Then, the perturbed magnetic field is evaluated with a set of linearized (in the perturbations), steady-state ($\partial /\partial t = 0$) two-fluid equations \cite{ferraro_2012}.
The total magnetic field is loaded into the gyrokinetic particle-in-cell (PIC) code XGC (X-point Gyrokinetic Code) for the calculation of neoclassical and turbulent plasma transport. 

XGC has been extensively benchmarked against other codes, e.g., the drift-kinetic code NEO \cite{hager2016_1,hager_2019_2} and the gyrokinetic codes GENE \cite{merlo_2018}, GTC \cite{holod_2013}, GEM \cite{hager_2017}, and EUTERPE \cite{cole_2019_2}.
A comprehensive summary of the edge-physics specific features of XGC can be found in \mycite{ku_2018}.

The 5D (3D configuration + 2D velocity space) gyrokinetic Boltzmann equation is solved in XGC with gyrokinetic ions and drift-kinetic electrons using the marker particle equations of motion \cite{ku_2018,rglittlejohn_1985,hahm_1988}
%
\begin{align} \label{eq:gk_equations}
 \frac{\partial f}{\partial t} &+ \dot{\boldsymbol{X}} \cdot \frac{\partial f}{\partial \boldsymbol{X}} 
               + \dot{v}_\parallel \cdot \frac{\partial f}{\partial v_\parallel} = S(f), \notag \\
 \dot{\boldsymbol{X}} &= \frac{1}{G} \left[ v_\parallel \left( \uvek{b} + \frac{\deltab}{B} \right)
                                           +\frac{m_s v_\parallel^2}{q_s B^2} \nabla \times \bnorm
                                           +\frac{1}{q_s B^2} \vek{B} \times \left( \mu \nabla B 
                                              - q_s \overline{\vek{E}} \right) \right], \notag \\
 \dot{v}_\parallel &= -\frac{1}{m_s G} \left( \bnorm + \frac{\deltab}{B} + \frac{m_s v_\parallel}{q_s B} \nabla \times \bnorm \right)
                                           \cdot \left( \mu \nabla B - q_s \overline{\vek{E}} \right), \notag \\
 G &= 1 + \frac{m_s v_\parallel}{q_s B} \bnorm \cdot \left( \nabla \times \bnorm \right),
\end{align}
where $f$ is the gyrokinetic distribution function, $\vek{X}=(R,\varphi,Z)$ is the gyrocenter position of marker particles in configuration space (using right-handed cylindrical coordinates with the $Z$-axis being the symmetry axis and $\varphi$ the toroidal angle), $\mu = m_s v_\perp^2/(2 B)$ is the magnetic moment, $v_\perp$ is the velocity perpendicular to the magnetic field, $v_\parallel$ is the velocity along the magnetic field, $m_s$ and $q_s$ are the mass and charge of species $s$, $B = |\vek{B}|$, $\bnorm = \vek{B}/B$, $\vek{B}$ is the axisymmetric equilibrium magnetic field, and $\deltab$ is the perturbed magnetic field imported from M3D-C1, which is of the order $\delta B/B \sim 10^{-3}$ or smaller.
$\delta\vek{B}$ includes both perpendicular and parallel perturbations as provided by M3D-C1.
The parallel particle flow is faithful to the perturbed magnetic field imported from M3D-C1, but the drift motions neglect higher order terms in $\deltab$.
The perturbed magnetic field imported from M3D-C1 $\deltab$ contains only the toroidal mode number $n=3$ as
in Ref. \onlinecite{hager_2019_1}, representing the toroidal mode number of the external RMP coil array.
Effects from side-band toroidal mode numbers and from neglecting the higher order drift terms in the present study result have not been studied and are not known.
All the poloidal mode structures, including the distortion from island overlapping, are retained as provided by M3D-C1.
For the gyrokinetic ions, $\overline{\vek{E}}$ is the gyroaveraged electric field, while for the assumed drift-kinetic electrons $\overline{\vek{E}} = \vek{E}$.

XGC  is a total-$f$ code in the sense that it calculates the total particle distribution function instead of calculating the perturbed part from a fixed Maxwellian background plasma  \cite{sku_2016}.
The full neoclassical physics drive is retained in $\dot{\vek{X}}\cdot \partial f/\partial\vek{X}$ without the Maxwellian background assumption.
Collisions \cite{yoon_2014,hager2016_2}, neutral particle recycling \cite{stotler_2017}, heat and torque sources \cite{ku_2018,hager_2019_1} are represented by the term $S(f)$ in Eq. \eqref{eq:gk_equations}.
Due to the high computational cost of XGC total-f turbulence simulations (see Sec. \ref{subsec:case_setup}), the perturbed magnetic field imported from M3D-C1 is not ramped up in this study.
That is, the RMP field is at full strength throughout the whole simulation in contrast to the slow, linear ramp-up used in the initial phase of the neoclassical transport simulations discussed in Ref. \mycite{hager_2019_1}.

The electrostatic potential is determined by the gyrokinetic Poisson equation \mycite{hahm_1988} using the Pad\'{e} approximation
\begin{equation}\label{eq:gk_poisson}
 \nabla_\perp \cdot \frac{n_e m_i}{q_i B^2} \nabla_\perp \phi = -\left( 1+ \nabla_\perp \cdot \rho_i^2 \nabla_\perp \right) \left(\overline{n}_i - n_e \right),
\end{equation}
where $\overline{n}_i$ is the gyroaveraged ion gyrocenter density, $n_e$ is the electron density, $\rho_i$ is the ion gyroradius, and $\nabla_\perp \equiv (\mytensor{I}-\bnorm\bnorm) \cdot \nabla$.
A logical sheath method \cite{parker_1993,ku_2018} is applied to determine the combined Debye and quasi-neutral sheath potential at the wall as a subgrid phenomenon.

\subsection{Test Case Setup}\label{subsec:case_setup}
\begin{figure}
 \centering
 \includegraphics{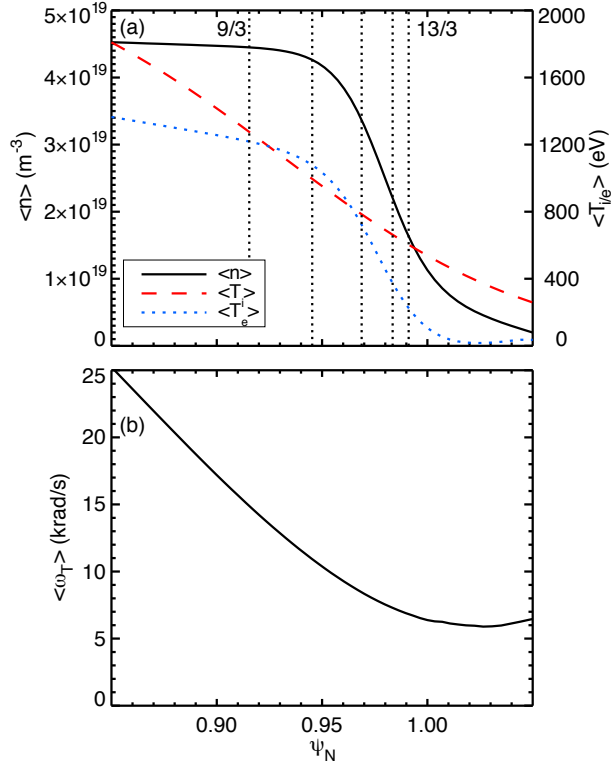}
 \caption{Initial conditions for XGCa simulations.
          (a) Density (solid line), ion (dashed line) and electron (dotted line) temperature, and the $(m,n)=(9/3)$, $(10,3)$, $(11,3)$, $(12,3)$, and $(13,3)$ rational surfaces (vertical dotted lines).
          (b) Toroidal rotation. $\omega_{T}>0$ implies counter-clockwise (co-current) rotation when looking from above.}
 \label{fig:initial_profiles}
\end{figure}

We investigate the same plasma as in Ref. \mycite{hager_2019_1}, a setup based on DIII-D H-mode discharge 157308 (see also Ref. \mycite{lyons_2018}).
The axisymmetric equilibrium magnetic field and the magnetic perturbation have been calculated in M3D-C1 based on an EFIT \cite{efit} reconstruction at 4.2 ms into the discharge.
At this point, the $n_{RMP}=3$, even parity RMP field is already at full strength with an I-coil current of 3.8 kA.
The density, temperature and toroidal rotation profiles used to initialize the simulation are shown in Fig. \ref{fig:initial_profiles}.
The initial plasma distribution function is constructed based on those profiles using shifted Maxwellians with density $n(\psi_N)=\overline{n}_i(\psi_N)=n_e(\psi_N)$, isotropic temperatures $T_{i/e}(\psi_N)$ and mean parallel flow $u = \vek{u}_T \cdot \uvek{b} = \omega(\psi_N) R B_T/B$, where $R$ is the major radius, $B_T$ is the toroidal magnetic field, and
 $\psi_N = (\psi-\psi_{ax})/(\psi_{sep}-\psi_{ax})$ is the axisymmetric poloidal magnetic flux normalized to its values on the magnetic axis ($\psi_{ax}$) and the separatrix ($\psi_{sep}$).
Thus, $\psi_N$ is zero at the magnetic axis and unity on the separatrix.
The initial electric field is zero because the right-hand side of Eq. \eqref{eq:gk_poisson} is zero with these initial conditions, but grows rapidly (in the form of ion polarization density) as the ion gyrocenters drift away from the electrons due to the large mass difference, and the particle distribution functions deviate from shifted Maxwellians.

%
\begin{figure}
 \centering
 \includegraphics{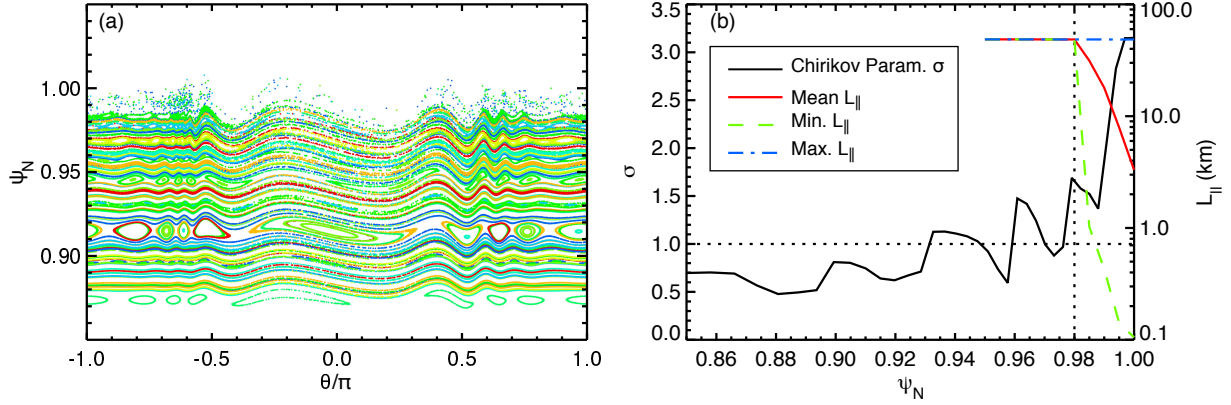}
 \caption{(a) Poincare plot of the total magnetic field (axisymmetric background and RMP field) at toroidal angle $\varphi=0$.
          (b) Average, minimal and maximal connection length to the material wall for field lines starting from the outer midplane at $0.95 \leq \psi_N \leq 1$ and from 100 uniformly spaced toroidal angles between $0$ and $2\pi$.
          Field line tracing is cut off after $\sim 50$ km (dashed-dotted blue line).
          The Chirikov parameter exceeds unity for $\psi_N \gtrsim 0.98$.
          [Reproduced from Ref. \onlinecite{hager_2019_1}, R. Hager \textit{et al.}, Nuclear Fusion \textbf{59}, 126009 (2019) \textcopyright2019 IOP Publishing Ltd]}
 \label{fig:delta_b_poincare_density}
\end{figure}

%
The perturbed magnetic field exhibits intact Kolmogorov-Arnold-Moser (KAM) surfaces in most of the edge pedestal volume [$\psi_N \lesssim 0.98$, Fig. \ref{fig:delta_b_poincare_density} (a)].
Only in the direct vicinity of the separatrix ($\psi_N \gtrsim 0.98$), the external RMP field makes the magnetic field weakly stochastic with a Chirikov criterion (as defined in Ref. \onlinecite{joseph_2012}) somewhat above unity as can be seen in Fig. \ref{fig:delta_b_poincare_density} (b).  
The weak and inhomogeneous stochasticity already implies that the idealized Rechester-Rosenbluth theory \cite{rechester_1978}, which assumes a fully stochastic system in the entire space, may not be valid here.
Although the sparsity of points in Fig. \ref{fig:delta_b_poincare_density} (a) does not allow to visualize them, there are many higher order Markov-tree islands up to the limit of M3D-C1 grid-resolution in a weakly stochastic toroidal system. 
This complicated process creates traps, or sticky orbits, and leads to sub-diffusive transport.
We refer the readers to Refs. \onlinecite{hudson_2006,meiss_2015,spizzo_2018} for further reading on how difficult it is for a perturbed toroidal magnetic field lines to satisfy the Rechester-Rosenbluth assumption.
The rapid variation of the Chirikov number across the pedestal, the presence of KAM surfaces, and the fact that the density pedestal width is comparable to the the ion banana orbit width $\Delta_b$ (see Fig. \ref{fig:initial_profiles}) are aspects of this initial state that are noteworthy and underscore the necessity of kinetic modeling of RMP-driven transport.

From the initial conditions described above, we performed two simulations with gyrokinetic deuterium ions and drift-kinetic electrons in order to assess the influence of the RMP field on coupled neoclassical and turbulent plasma transport.
The first simulation is run without RMP field, i.e. using only the axisymmetric equilibrium magnetic field.
The second simulation is run with full-strength RMP field starting from $t=0$.
The simulated physics time is long enough to extract transport fluxes and effective transport coefficients, but not long enough to yield noticeably different plasma profiles in the two simulations.  

Experimental level core heat and torque sources are included.
A total of 5.95 MW of heating power is applied at normalized poloidal flux $0.1 \leq \psi_N \leq 0.93$, split evenly among ions and electrons.
In addition, a constant electron heat sink of 3 MW is applied in the edge layer at $0.995 \leq \psi_N \leq 1.1$ as a simplified proxy for radiative power loss.
A co-current torque source of 1.7 Nm is applied at $0.12 \leq \psi_N \leq 0.93$ to account for the neutral beam torque.
The particle recycling rate, i.e. the fraction of particles hitting the material wall that are reintroduced as neutrals, is set to 0.95 in both simulations to account for the high vacuum pump rate in the RMP experiments.
In XGC simulations without RMPs, we usually set the recycling rate to 0.99.

In order to save computational resources, the simulation is stopped at time $t=0.2$ ms.
This time is much longer than the neoclassical ion transit time scale $\tau_{t,i}=q_{95} R_0/v_T\approx 0.04$ ms \cite{hager_2019_1} (saturation of the mean radial electric field) and the drift-wave time scale $L_p/(k_\perp \rho_i v_T) < 10^{-2}$ ms at $\psi_N=0.95$ (saturation of turbulence), where $q_{95}$ is the safety factor at $\psi_N=0.95$, $R_0$ is the major radius on the magnetic axis, $k_\perp$ is the perpendicular wave number, $L_p$ is the pedestal gradient scale length, and $v_T=(2 k_B T_i/m_i)^{1/2}$.
However, the simulation time of 0.2 ms corresponds to only about one third of the ion-ion collision time $\tau_{c,i}$ at $\psi_N=0.95$, which is not long enough to obtain saturated neoclassical ion thermal conductivity.
Since ion thermal transport from neoclassical physics may not be negligible compared to that from turbulence, discussion on the ion thermal conductivity is excluded from this article.
The geodesic acoustic modes (GAMs) resulting from the relaxation of the initial plasma state tend to be strongly Landau-damped because they are quickly sheared to higher radial wavenumbers $k_r$ due to the temperature dependence of the GAM frequency \cite{qiu_2008,xu_xq_2008}.

As is extensively discussed in Ref. \onlinecite{hager_2019_1}, this simulation time scale (which is much longer than the ion transit time scale) is sufficient for the saturation of the mean radial electric field in a pedestal with steep gradients.
Unlike in core plasma with mild radial gradients,  where collisional damping dominates the saturation of the mean radial electric field \cite{novakovskii_1997}, the fast radial ion excursion across the steep plasma gradient dominates the saturation time-scale in an edge pedestal.
Saturation of the radial electric field is accompanied by the saturation of the ambipolar radial particle flux, as is well explained in Ref. \onlinecite{hirshman_1978}.
We refer the reader to Fig. 3 (a) of Ref. \onlinecite{hager_2019_1} for the early saturation of the neoclassical particle flux at $\sim 0.1\tau_{i,c}$ in the steep pedestal.
Since the neoclassical electron heat-transport is negligible, the electron thermal conductivity reaches the quasi-steady state at the fast turbulence saturation time scale in the edge pedestal $\psi_N \gtrsim 0.95$, as shown in Sec. \ref{sec:results}.

In order to make the computational cost manageable, we make use of the $n_{RMP}=3$ toroidal periodicity of the RMP coils and simulate only a 1/3 wedge of the full torus with periodic boundary condition in the toroidal direction and 16 poloidal planes in the 1/3 wedge.
The configuration space mesh for this case has a radial and poloidal resolution of approximately 1 mm ($\approx \rho_i/2$) in the pedestal region $0.9 < \psi_N < 1$, with somewhat coarser resolution in the central core and the far scrape-off layer.
Note that due to the field-alignment of the XGC mesh \cite{adams_ku_2009,fzhang_2015}, the resolution of high poloidal mode numbers $m$ with 1 mm poloidal grid-spacing allows for resolving much higher toroidal mode numbers $n$ (with $m \simeq n q$) than in a conventional mesh with 16 poloidal planes in the 1/3 wedge.
The number of configuration space vertices is 181,162 per $\varphi=\text{constant}$-plane, i.e., the total mesh size is close to 3 million vertices in the 1/3 wedge.
A total of 24.1 billion marker particles per particle species is used corresponding to roughly 8,300 particles per mesh vertex and species.
Each simulation was run using 2,048 compute nodes ($\approx 45$\%) of the $\approx$12 PFLOPS Theta computer at Argonne Leadership Computing Facility (ALCF) for approximately 7 days.

%
\section{Results}\label{sec:results}

Our simulation results are presented here in two parts.
First, we discuss the general properties of the turbulence activity observed in the simulations with and without RMPs.
In the second part, we investigate the relation between turbulence activity and transport fluxes.

\subsection{Classification of turbulent fluctuations}\label{subsec:turbulence}

\begin{figure}
 \centering
 \includegraphics{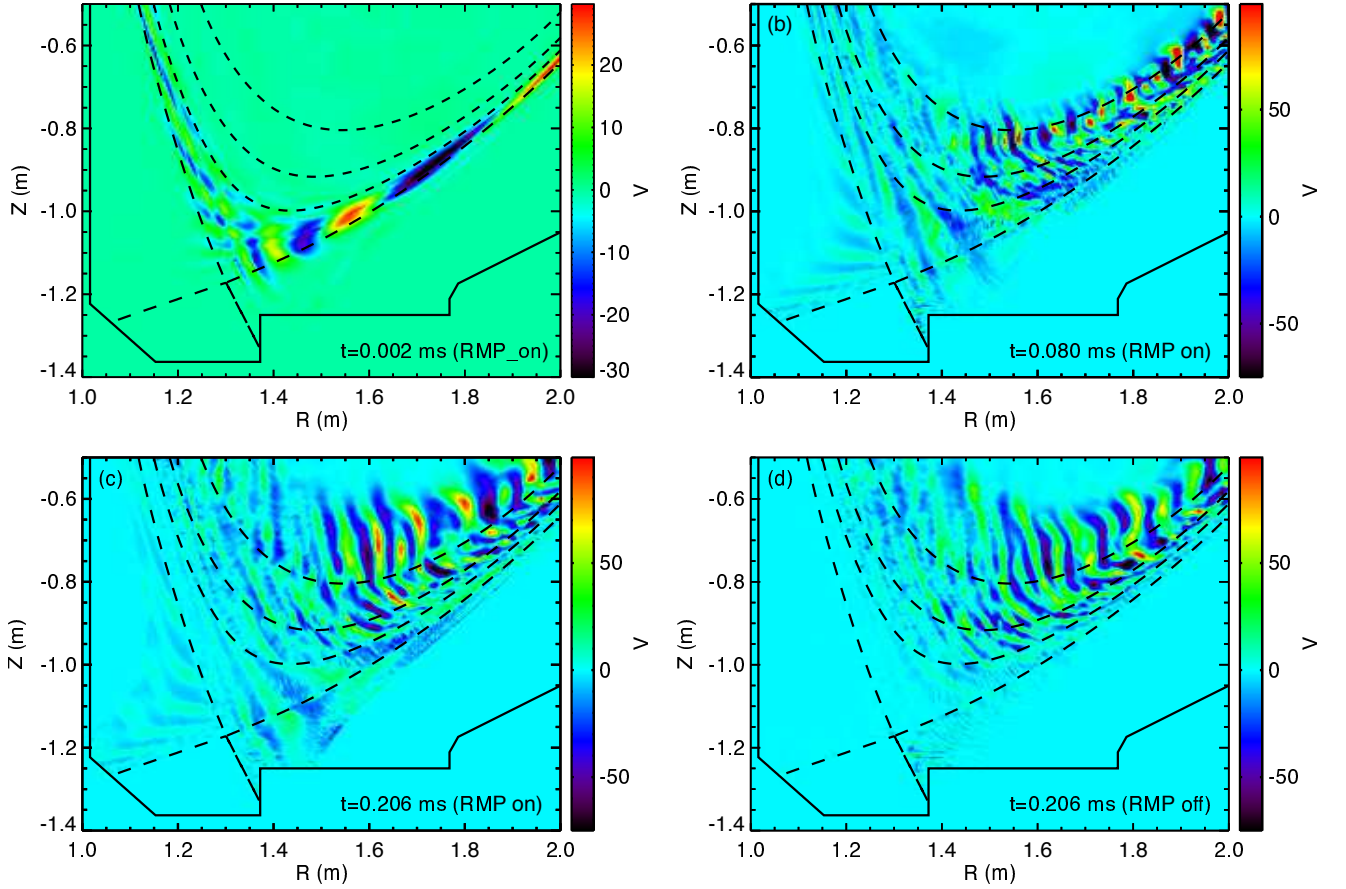}
 \caption{Non-axisymmetric potential perturbation $\delta\phi_t$ on the radial-poloidal plane $\varphi=0$ in Volt (V):
          (a) $t=0.002$ ms, with RMP field;
          (b) $t=0.080$ ms, with RMP field; 
          (c) $t=0.206$ ms, with RMP field;
          (d) $t=0.206$ ms, without RMP field.
          The solid line marks the material wall, the dashed lines mark the flux-surfaces $\psi_N=0.8$, 0.9, 0.95 and 1.}
 \label{fig:dphi_snapshots}
\end{figure}

Simply from visual inspection of the non-axisymmetric electrostatic potential $\delta \phi_t = \phi - \langle \phi \rangle_T$, where $\langle \dots \rangle_T = 1/(2\pi) \oint \dots \mathrm{d}\varphi$ is the toroidal average, two RMP-induced effects become apparent.
Snapshots of $\delta \phi_t$ at $t=2.0\cdot 10^{-3}$, 0.08 and 0.206 ms are shown in Fig. \ref{fig:dphi_snapshots} (a)-(c) for the case with RMP field and at $t=0.206$ ms in Fig. \ref{fig:dphi_snapshots} (d) for the case without RMP field.
While there is no visual activity of $\delta \phi_t$ at $t=2.0\cdot 10^{-3}$ ms without RMPs (not shown), there is a characteristic $n=3$ response in $\delta \phi_t$ that is most pronounced in the edge pedestal in Fig. \ref{fig:dphi_snapshots} (a).
This $n=3$ response, that is discussed in detail in Ref. \onlinecite{hager_2019_1}, is due to the equilibration of the electrostatic potential on the perturbed flux-surfaces.
It is expected to form on the electron toroidal transit time scale (1 $\mu$s at $\psi_N=0.97$) and is naturally strongest where the radial electric field (or the pressure gradient) is largest.
Figure \ref{fig:dphi_snapshots} (b) shows micro-instabilities growing at relatively small scales in the edge pedestal with RMPs while the core plasma is still quiescent.
At later times, instabilities start to grow in the core plasma as well at larger scale than in the edge as illustrated in \ref{fig:dphi_snapshots} (c).
At $t=0.206$ ms, the turbulent structures in $\delta \phi_t$ look rather similar in both cases [with RMPs in Fig. \ref{fig:dphi_snapshots} (c) and without RMPs in Fig. \ref{fig:dphi_snapshots} (d)] except in the scrape-off layer (SOL), where the simulation with RMP field exhibits stronger electrostatic potential perturbations than those without RMP field.
Lobe structures can also be seen around the separatrix with RMPs in Fig. \ref{fig:dphi_snapshots} (c).

\begin{figure}
 \centering
 \includegraphics{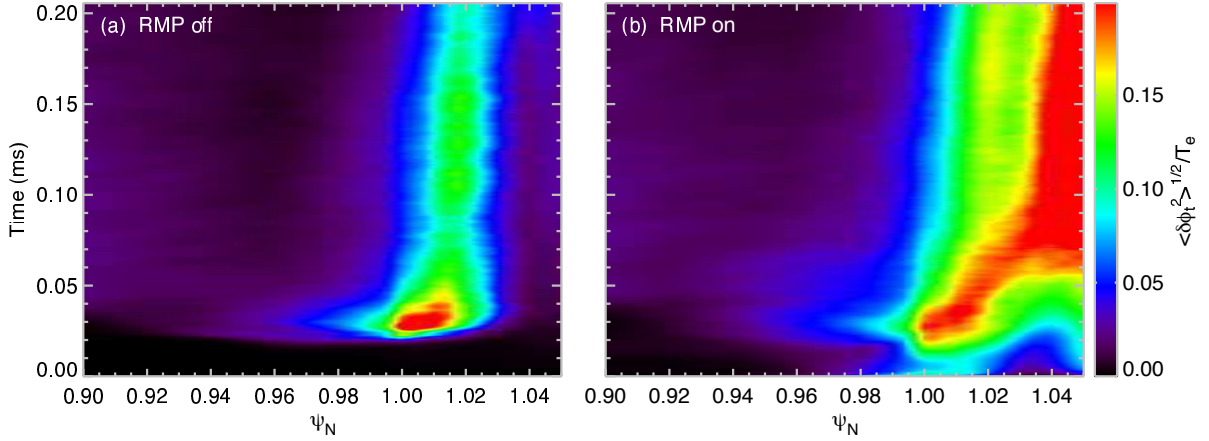}
 \caption{Normalized RMS amplitude of the electrostatic potential perturbation $e \langle\delta\phi_t^2\rangle^{1/2}/(k_B \langle T_e\rangle$ vs. normalized poloidal flux $\psi_N$ and time: (a) without RMP field; (b) with RMP field.}
 \label{fig:dphi_amplitude_psi_t}
\end{figure}

These visual findings are confirmed by calculating the flux-surface averaged normalized RMS amplitude of the non-axisymmetric potential perturbations $e \langle ( \delta \phi_t)^2 \rangle^{1/2}/\langle k_B T_e \rangle)$, where $\langle \dots \rangle$ indicates flux-surface averaging (see Fig. \ref{fig:dphi_amplitude_psi_t}).
Notice here that Fig. \ref{fig:dphi_amplitude_psi_t} shows a stronger amplitude in the scrape-off layer than Fig. \ref{fig:dphi_snapshots} since the potential perturbation is normalized to the local flux-surface-averaged electron temperature $\langle T_e \rangle$.
Figure \ref{fig:dphi_amplitude_psi_t} also demonstrates that 
the transient turbulence behavior settles down to a quasi-steady phase after 0.05 ms without RMPs and after 0.1 ms with RMPs.
After these times, the turbulence is slowly evolving because the background density and temperatures are evolving, so that the radial transport properties can be measured. 
We note here that a significant part of the potential perturbation in the scrape-off layer (SOL) is identified to be low-$k_\parallel$ modes with toroidal mode number $n=3$.

\begin{figure}
 \centering
 \includegraphics{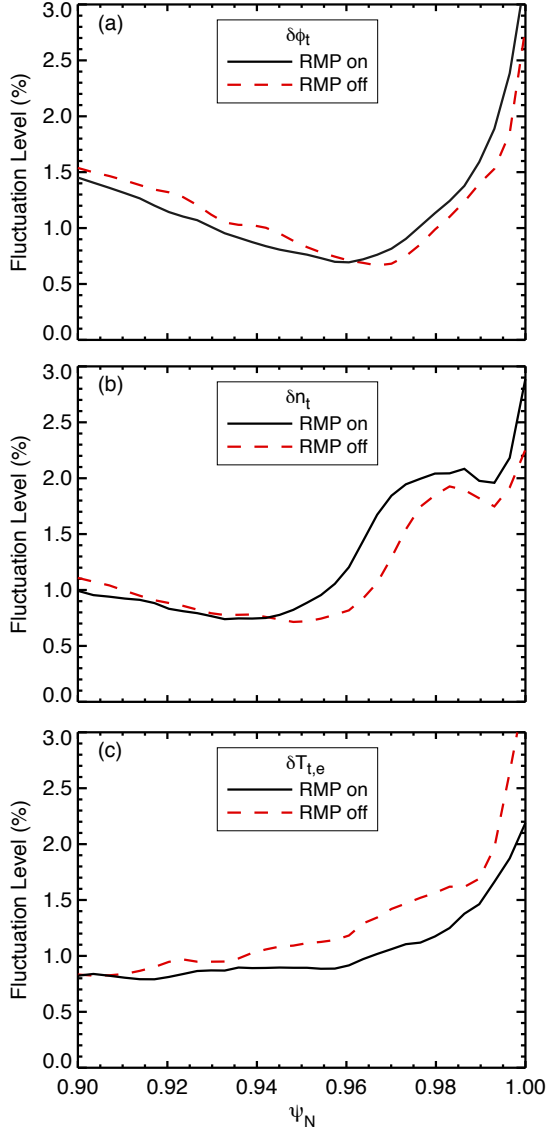}
 \caption{Normalized RMS amplitudes of (a) the electrostatic potential, (b) density and (c) electron temperature fluctuations $e \langle\delta\phi_t^2\rangle^{1/2}/(k_B \langle T_e\rangle$, $\langle \delta n_t^2\rangle^{1/2}/\langle n \langle$, and $\langle \delta T_{t,e}^2 \rangle^{1/2}/\langle T_e \rangle$ time-averaged over $0.191 \text{ ms} \leq t \leq 0.206$ ms.}
 \label{fig:dphi_amplitude_n3}
\end{figure}

Pure turbulence features without the interference from the $n=3$ background perturbation can be inspected by filtering out from the density, temperature and potential fluctuations the low $k_\parallel$ modes near $m=3q$, e.g., $3q - 7 \leq m \leq 3q+7$.
The results, time-averaged between $0.191 \text{ ms} \leq t \leq 0.206$ ms, are shown in Fig. \ref{fig:dphi_amplitude_n3}.
From Fig. \ref{fig:dphi_amplitude_n3}, interesting observations can be made about the filtered fluctuations:
i) The potential perturbation $e \langle ( \delta \phi_t)^2 \rangle^{1/2}/\langle k_B T_e \rangle$ increases with RMPs at $\psi_N >0.96$ and decreases at $\psi_N<0.96$, although those changes of about 10\% are relatively small meaning that other quantities need to be examined to judge the effect of the RMPs on the turbulent transport;
ii) the density perturbation $\langle ( \delta n_t)^2 \rangle^{1/2}/\langle n \rangle)$ increases more strongly than the potential perturbation, by $\sim$40\% at $\psi_N=0.97$;
and iii) the electron temperature perturbation $\langle ( \delta T_{e,t})^2 \rangle^{1/2}/\langle T_e \rangle)$ exhibits reduced amplitude (by $\sim$25\% at $\psi_N=0.97$).
It will be shown later that the increased density fluctuations under RMPs are correlated with the enhanced particle transport.
%

In order to classify the instabilities behind the turbulence activity in our simulation, we mapped the electrostatic potential perturbation to an observation window around the outer midplane ($0.94 \leq \psi_N \leq 1.0$ and $-0.3 \lesssim Z \lesssim 0.3$) using a rectangular uniform grid in $\psi_N$ and $l_\theta$, where $l_\theta$ is the arclength along a flux-surface at constant toroidal angle $\varphi$.
The positive poloidal direction 
corresponds to the electron diamagnetic direction.
In the time window $0.128 \leq t \leq 0.206$ ms, we evaluated the fluctuation spectrum in $k_\theta \rho_i$ ($\rho_i$ is the ion gyroradius) and frequency $\omega$, as well as the time and flux-surface averaged poloidal $\exb$ velocity $\langle v_{E,\theta}\rangle = \langle B^{-2} (\vek{E}\times\vek{B}) \cdot \hat{\vek{\theta}} \rangle$, where $\hat{\vek{\theta}}$ is the unit vector in the poloidal direction.
In the following, negative velocity corresponds to motion in the ion diamagnetic direction (downward on the outer midplane).
The $\exb$ velocity and the phase velocity of the turbulent modes in the laboratory frame indicate whether an unstable mode is an ion or an electron drift mode.

\begin{figure}
 \centering
 \includegraphics{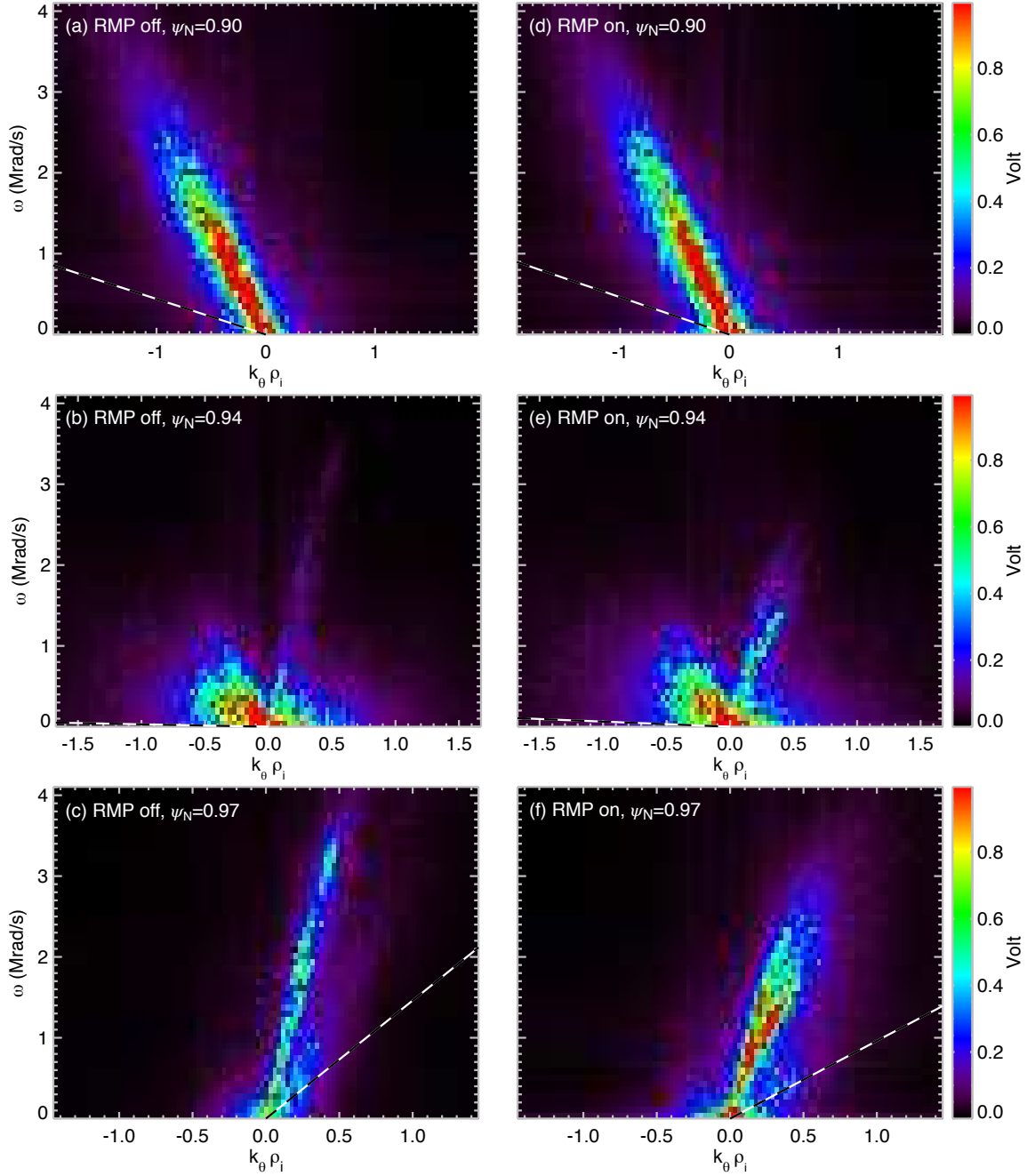}
 \caption{Spectra of the electrostatic potential perturbation $\delta\phi_t$ in poloidal wavenumber $k_\theta \rho_i$ and frequency $\omega$ on flux-surfaces $\psi_N=0.9$ [(a) and (d)], $\psi_N=0.94$ [(b) and (e)], and $\psi_N=0.97$ [(c) and (f)]. Plots (a)-(c) show results without and plots (d)-(f) results with RMP field.}
 \label{fig:dphi_spectra}
\end{figure}

A series of spectra at $\psi_N=0.9$, 0.94, and 0.97 is shown in Fig. \ref{fig:dphi_spectra} (a)-(c) for the case without RMP field and (d)-(f) for the case with RMP field.
At $\psi_N=0.9$ (pedestal top), we find a well defined mode with negative phase velocity $v_{ph} \approx 6 \langle v_{E,\theta} \rangle$ and similar amplitude with and without RMP field [Figs. \ref{fig:dphi_spectra} (a) and (d)].
This mode is clearly an ion drift mode because $v_{ph} < \langle v_{E,\theta} \rangle < 0$ implies motion in the ion diamagnetic drift direction in the moving $\exb$ frame.
The spectra at $\psi_N=0.94$ [Figs. \ref{fig:dphi_spectra} (b) and (e)] indicate the presence of two different modes, one moving in the ion and one moving in the electron diamagnetic direction in the $\exb$ frame.
The electron mode has higher amplitude with the RMP field switched on [Fig. \ref{fig:dphi_spectra} (e)].
Finally, the spectra at $\psi_N=0.97$ [Figs. \ref{fig:dphi_spectra} (c) and (f)], which is in the steepest part of the edge pedestal, exhibit only the electron mode while the ion mode disappears on this flux-surface.
On all three flux-surfaces discussed here, the mean poloidal scale of the turbulence is $k_\theta \rho_i = 2\pi \rho_i/\lambda_\theta \approx 0.25$, i.e. $\lambda_\theta \approx 25 \rho_i$.
Since the present XGC simulations use the electrostatic approximation, these observations suggest that ion temperature gradient-driven modes (ITG modes) are dominant at the pedestal top, trapped electron modes (TEMs) are dominant at the steepest part of the pedestal, and a mixture of ITG modes and TEMs in between.
The turbulence modes enhanced by RMPs can be identified to be the trapped electron modes at lower frequency as can be noticed by comparing Figs. \ref{fig:dphi_spectra} (c) and (f).
With RMPs, TEMs are much stronger at lower frequency with weaker high frequency tail.
Without RMPs, TEMs are weaker at lower frequency but with a stretched tail to higher frequencies.
This is an indication of an $\exb$ shearing effect that will be discussed later.
\subsection{Neoclassical and turbulent transport}\label{subsec:transport}
The turbulence properties described in Sec. \ref{subsec:turbulence}, especially the increased density fluctuation amplitude and the stronger modes moving in the electron diamagnetic direction at $\psi_N \gtrsim 0.94$ with RMPs, already hint at potential consequences for radial transport fluxes.
In the following, we analyze those transport fluxes in some detail.

The definitions of flux densities in this article follows Eq. (4) in Ref. \onlinecite{hager_2019_1}.
Particle flux density $\Gamma$, energy flux density $Q$, and heat flux density $q$ are defined as
\begin{align}\label{eq:flux_definition}
  \Gamma &= \int \gamma^{-1} \nabla \psi \cdot \left( \boldsymbol{v}_E + \boldsymbol{v}_D   
                +v_\parallel \frac{\delta\boldsymbol{B}}{B} \right) f \mathrm{d}^3v, \notag \\
  Q &= \int \gamma^{-1} \nabla \psi \cdot \left( \boldsymbol{v}_E + \boldsymbol{v}_D   
                +v_\parallel \frac{\delta\boldsymbol{B}}{B} \right)\,
                \frac{m}{2}\left( v_\perp^2 + v_\parallel^2 \right) f \mathrm{d}^3v, \notag \\
  q &=  Q - \frac{5}{3} \gamma^{-1} \nabla \psi \cdot \boldsymbol{u} 
                        \int \frac{m}{2}\left( v_\perp^2 + v_\parallel^2 \right) f \mathrm{d}^3v
                     = Q - \frac{5}{2} k_B T \Gamma,
\end{align}
where $\boldsymbol{v}_E$ is the $\exb$-drift velocity, $\vek{v}_D$ is the magnetic drift velocity ($\nabla B$ and curvature), 
$\gamma=\langle |\nabla \psi|^2 \rangle^{1/2}$, and $\gamma^{-1} \nabla \psi \cdot \boldsymbol{u} = \gamma^{-1} \nabla \psi \cdot \boldsymbol{\Gamma}/n$ is the radial bulk velocity of the plasma.
For the ion species these flux densities are gyrocenter flux densities, i.e., they do not include polarization effects.

Unlike in an axisymmetric tokamak where any non-axisymmetric perturbations can be considered to be from turbulence, in a toroidal magnetic confinement configuration with non-axisymmetric equilibrium magnetic field perturbations, it is difficult to make a clear distinction between the neoclassical and turbulent transport fluxes because of the overlap of the mode numbers in the equilibrium magnetic field and the mode numbers in the turbulent perturbation.
If the combined turbulent-neoclassical simulation could be performed over a much longer time-period than the slowest turbulent mode in the system, then one could perform a long-time average and define the time-mean fluxes as neoclassical.
However, this would would require an unknown amount of computing resources, and would still miss the possible turbulence effect on the mean non-axisymmetric perturbation of physics quantities.
In this work, we define the neoclassical neoclassical fluxes to be those stemming from the axisymmetric parts of the drift velocities $\langle\boldsymbol{v}_D\rangle_T+\langle \boldsymbol{v}_E \rangle_T$ and $\langle f  \rangle_T$, and from those explicitly resulting from the parallel particle motion along the perturbed magnetic field $v_\parallel \delta\boldsymbol{B}/B$ in interaction with $\tilde{f}$, where $\tilde{f} = f-\langle f \rangle_T$ is the non-axisymmetric part of the distribution function.
In this context, there is a possibility that $\tilde{f}$ could include some effect from turbulent fluctuations with the same toroidal periodicity as $\delta\boldsymbol{B}$.
The turbulent fluxes are defined to be from the non-axisymmetric $\exb$ perturbations $\tilde{\boldsymbol{v}}_E = \boldsymbol{v}_E-\langle \boldsymbol{v}_E \rangle_T$ in interaction with the non-axisymmetric $\tilde{f}$.
Again, in this definition of the turbulent fluxes, some neoclassical fluxes, as defined in \onlinecite{hager_2019_1}, will be included through the non-axisymmetric $\delta\boldsymbol{B}$-driven $\tilde{\boldsymbol{v}}_E$ and $\tilde{f}$.
Thus, even though these cross-effects are expected to be small, a direct comparison of the present neoclassical fluxes and effective diffusion coefficients against those presented in \onlinecite{hager_2019_1} will not be highly accurate.
The effective transport coefficients are defined analogous to Eq. (5) in Ref. \onlinecite{hager_2019_1}.
In the following discussion, the neoclassical flux densities and the effective transport coefficients are marked with a subscript ``$neo$'' while turbulent flux densities and the effective transport coefficients are marked with a subscript ``$turb$''.

\begin{figure}
 \centering
 \includegraphics{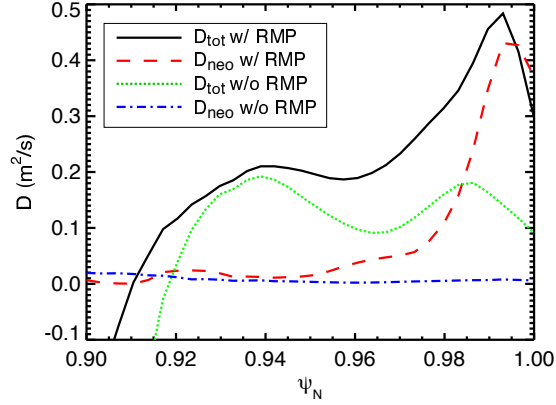}
 \caption{Time-averaged ($0.191 \text{ ms} \leq t \leq 0.206$ ms) effective particle diffusivity: Total particle diffusivity $D_{tot}=D_{turb}+D_{neo}$ with (solid black line) and without (dotted green line) RMP field shown together with the neoclassical particle diffusivity $D_{neo}$ with (dashed red line) and without (dashed-dotted blue line) RMP field.}
 \label{fig:particle_diff}
\end{figure}

The XGC-evaluated effective particle diffusivities are plotted in Fig. \ref{fig:particle_diff}.
The time-averaged ($0.191 \text{ ms} \leq t \leq 0.206$ ms) total particle diffusivity $D_{tot}=D_{neo}+D_{turb}$ [solid black line in Fig. \ref{fig:particle_diff}] is significantly enhanced by the RMP field in the steep pedestal region $\psi_N \gtrsim 0.94$.
The RMP-driven neoclassical contribution $D_{neo}$ [dashed red line in Fig. \ref{fig:particle_diff} (a)] exceeds the turbulence contribution $D_{turb}$ in the 
 weakly 
stochastic layer around the separatrix at $\psi_N\gtrsim 0.985$.
But at $0.94 \lesssim \psi_N <0.985$, majority of the enhanced particle diffusivity is due to turbulent transport.
In the absence of an RMP field, the neoclassical contribution (blue dashed-dotted line) to the particle flux is negligible compared to the turbulence contribution.

\begin{figure}
 \centering
 \includegraphics{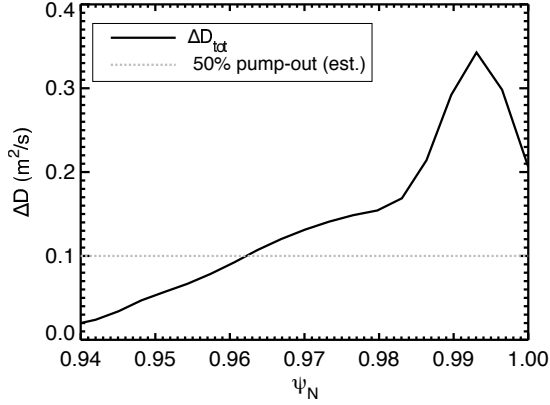}
 \caption{Difference between the total (turbulent and neoclassical) particle diffusivities with and without RMP field.
         A rough estimate for the additional particle diffusivity that is required for experimentally relevant (50\%) density pump-out in less than 100ms is $\Delta D \gtrsim 0.1 \,\text{m}^2 \text{s}^{-1}$.}
 \label{fig:delta_d_rmp}
\end{figure}

A simple estimate of the density pump-out rate that can be expected from this increased particle diffusivity is obtained using the following assumptions, as described in Ref. \citenum{hager_2019_1}: i) The particle diffusivity in the reference simulation without RMP field maintains the plasma in steady state for constant particle source, and $D$ must increase due to the RMP field for particle pump-out with a given radial density gradient;
and ii) 50\% of the total plasma mass is pumped out within less than 100 ms due to the RMP field. 
Under these assumptions, we find that the RMP-induced additional particle diffusivity $\Delta D_{RMP}$ required for such a density pump-out is approximately $\gtrsim 0.1 \text{m}^2/\text{s}$ in the pedestal region.
This estimate is of the same order as the experimental-data based estimate given in Ref. \onlinecite{evans_2006}.
A more rigorous approach to find the new steady state plasma profiles with RMP field would be to use the XGC-computed particle diffusivity in a transport code.
This has not been done for this study, but is being pursued.
The comparison between the simulations with and without RMP field yields $\Delta D_{RMP}=D_{RMP}-D_{ref}>0.1$ at $\psi_N \gtrsim 0.96$ (Fig. \ref{fig:delta_d_rmp}).
This is an important result because it suggests that the experimentally observed particle pump-out in the entire pedestal slope region can only be explained by combining the neoclassical and turbulent physics. 
%

\begin{figure}
 \centering
 \includegraphics{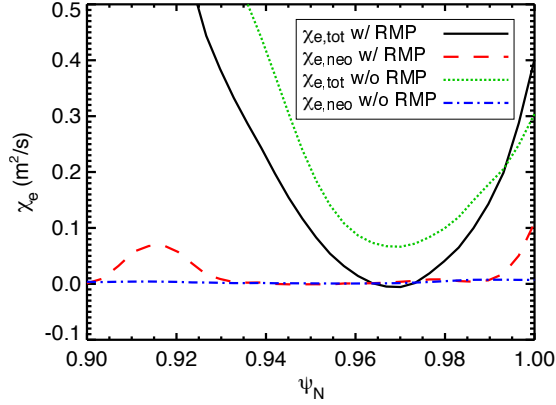}
 \caption{Time-averaged ($0.191 \text{ ms} \leq t \leq 0.206$ ms) effective electron heat conductivity: total electron heat conductivity $\chi_{e,tot}=\chi_{e,turb}+\chi_{e,neo}$ with (solid black line) and without (dotted green line) RMP field; and neoclassical electron heat conductivity $\chi_{e,neo}$ with (dashed red line) and without (dashed-dotted blue line) RMP field.}
 \label{fig:chi_e_comparison}
\end{figure}

Another interesting effect from RMP fields besides the density pump-out is the preservation or steepening of the electron temperature gradient in the steep pedestal region \cite{evans_2006}.
Our previous study of neoclassical transport with RMP fields could not reach a firm conclusion on electron heat transport \cite{hager_2019_1} because it omitted turbulent heat flux, which is much larger than neoclassical electron heat flux.
In the present study, we find that, indeed, $\chi_{e,turb} \gg \chi_{e,neo}$ and that the electron heat flux in the steepest part of the electron temperature pedestal is suppressed by the RMP field (even reaching a negligibly small level at $0.96<\psi_N < 0.98$; see Fig. \ref{fig:chi_e_comparison}).
The neoclassical electron heat conductivity is negligible compared to the turbulent transport with and without RMP field.
Thus, our simulation results are qualitatively consistent with experimental observation.
It is noted here again that the insignificant increase of the effective neoclassical $\chi_{e,neo}$ in the pedestal-foot region (red dashed line in Fig. \ref{fig:chi_e_comparison}), where the field-lines appear to be stochastic in Fig. \ref{fig:delta_b_poincare_density} (a) is because of the weakness in the stochasticity level as seen in Fig. \ref{fig:delta_b_poincare_density} (b), as explained earlier.
%

\begin{figure}
 \centering
 \includegraphics{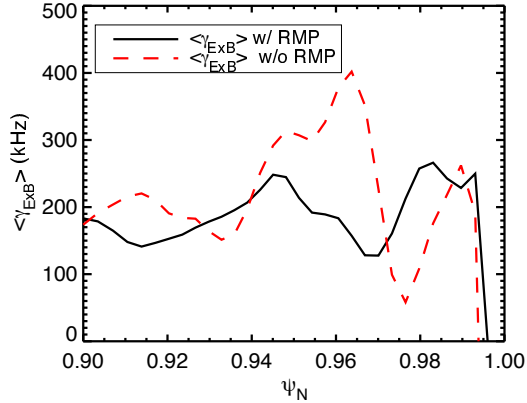}
 \caption{Time and flux-surface averaged $\exb$ shearing rate $\langle \gamma_{E\times B}\rangle$ without (solid black line) and with (dashed red line) RMP field.}
 \label{fig:shearing_rate}
\end{figure}

The RMP field can cause significant changes in the $\exb$ shearing rate \cite{hager_2019_1,schmitz_2018} $\langle\gamma_{E\times B}\rangle=\langle|\nabla \psi_N|\rangle\, \partial \langle E_r (B_T/B^2)\rangle/\partial \psi_N$ and possibly explain the changes in the electron-directional turbulence (see Fig. \ref{fig:dphi_spectra}) and the radial transport \cite{taimourzadeh_2019}.
Comparing the time averaged ($0.191 \text{ ms} \leq t \leq 0.206$ ms) shearing rate profiles $\langle \gamma_{E\times B}\rangle$ between the reference and RMP simulations (Fig. \ref{fig:shearing_rate}), one finds that $\langle \gamma_{E\times B}\rangle$ in the RMP simulation is reduced by up to about 50\% at $0.94 \lesssim \psi_N \lesssim 0.97$, but increased at $0.97 \lesssim \psi_N \lesssim 0.99$.
While the overall change of $\langle \gamma_{E\times B}\rangle$ in the pedestal due to the RMP field is significant and largely consistent with the predator-prey scenario \cite{biglari_shearflow_and_turbulence} that a lower mean shearing rate yields higher transport fluxes.
Some details do not completely agree at the local level, though:
We observe that the $\exb$ shearing rate is increased by the RMP field at $0.97 \lesssim \psi_N \lesssim 0.99$, but the local particle flux is still increased.
It is conjectured that the nonlocal effect due to the long wavelength nature of our trapped electron modes ($k_\perp\sim 0.25$) may be responsible for this local inconsistency.

In order to obtain a more detailed picture of the difference in turbulent transport between the simulations with and without RMPs, we examine the cross-spectra between the radial $\exb$ velocity $v_{E,r}$ and the non-axisymmetric density and temperature perturbations $\delta n_t = n - \langle n \rangle_T$ and $\delta T_{e,t} = T_e - \langle T_e \rangle_T$.
We focus on the turbulent $\exb$ transport fluxes, because the cause for the enhanced neoclassical transport with RMP field at $\psi_N \gtrsim 0.98$ has already been investigated in Ref. \onlinecite{hager_2019_1}.
Here, the cross-spectrum of two quantities $A(\psi_N,\theta^\ast,\varphi)$ and $B(\psi_N,\theta^\ast,\varphi)$ is defined as
\begin{equation}\label{eq:cross_spectrum_def}
S_{AB} = \langle \tilde{A}(\psi_N,m,\varphi) \tilde{B}^\ast(\psi_N,m,\varphi) \rangle_T
\end{equation}
with the straight field-line angle $\theta^\ast$ (see Ref. \onlinecite{hager_2019_1}), the poloidal mode number $m$, and $\tilde{\dots}$ indicating the poloidal Fourier transform.
That is, $S_{AB}$ is the poloidal Fourier transform of the cross-correlation of $A$ and $B$.
The cross-power is $P_{AB}=|S_{AB}|$, and the cross-phase is $\delta\zeta_{AB} = \arctan[\Im (S_{AB})/\Re (S_{AB}) ]$.
From cross-power and cross-phase, one can reconstruct the contributions to the flux-surface averaged radial particle, energy, and heat flux densities as
\begin{align}\label{eq:flux_per_mode}
  \Gamma_t(\psi_N,m) &= \alpha(\psi_N) P_{v n} \cos(\delta\zeta_{v n}), \notag\\
  Q_t(\psi_N,m) &= \alpha(\psi_N)\, 3 k_B/2 [\langle n \rangle P_{vT} \cos(\delta\zeta_{vT}) + \langle T \rangle P_{v n} \cos(\delta\zeta_{n T})], \notag \\
  q_t(\psi_N,m) &= Q_t - 5/2 k_B \langle T \rangle \Gamma_t,
\end{align}
where $\alpha(\psi_N)$ is a factor that accounts for the difference between the definitions of the cross-spectral density $S_{AB}$ and the flux-surface average
\begin{equation}\label{eq:flux_avg_definition}
  \langle A \rangle = \oint A R^2 \mathrm{d}\theta^\ast \mathrm{d}\varphi / \oint R^2 \mathrm{d}\theta^\ast \mathrm{d}\varphi.
\end{equation}
The factor $R^2$ in the numerator of the flux-surface average would usually require evaluation of the convolution of $\tilde{R^2}$ and $S_{AB}$ in Fourier space.
Since $\tilde{R^2}(\psi_N,k_\theta)$ decays rapidly at $|k_\theta| >0$, we approximate this convolution by using
\begin{equation}\label{eq:flux_avg_approximation}
  \tilde{R^2}(\psi_N,k_\theta) \approx \delta(k_\theta)\, \int |\tilde{R^2}(\psi_N,k_\theta)| \mathrm{d} k_\theta,
\end{equation}
so that 
\begin{equation}
  \alpha(\psi_N) = \int |\tilde{R^2}(\psi_N,k_\theta)| \mathrm{d} k_\theta / \oint R^2(\psi_N,\theta^\ast) \mathrm{d}\theta^\ast \mathrm{d}\varphi.
\end{equation}

\begin{figure}
 \centering
 \includegraphics{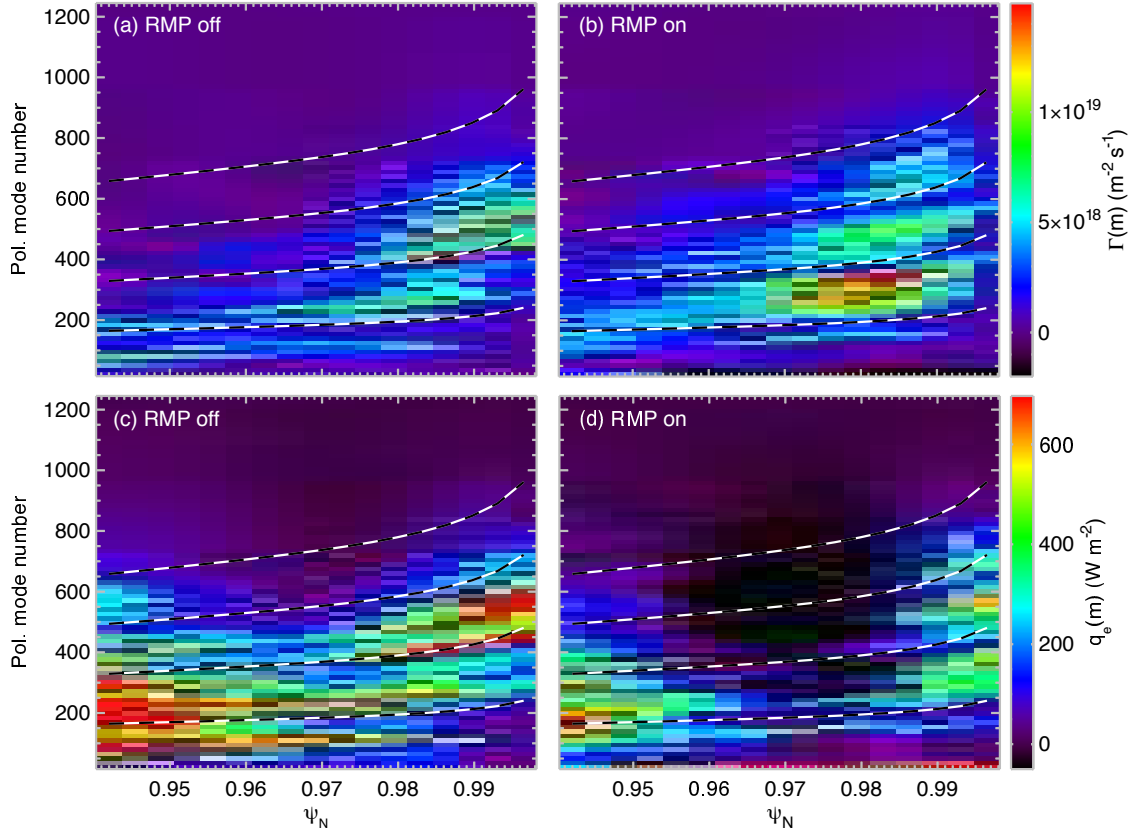}
 \caption{Particle and electron heat flux split into contributions from different poloidal mode numbers $m$.
          The dashed lines indicate $m=n q$ for $n=50$, 100, 150, and 200.
          Subfigures (a) and (b) show the particle flux density with and without RMPs, respectively.
          Subfigures (c) and (d) show the electron heat flux density with and without RMPs, respectively.}
 \label{fig:flux_per_mode}
\end{figure}

Figures \ref{fig:flux_per_mode} (a) and (b) show the particle flux density $\Gamma_t(\psi_N,m)$ with and without RMP field, respectively.
The dashed lines indicate the poloidal mode numbers $m=nq$ for $n=50$, 100, 150, and 200.
Most of the RMP-driven increase in the turbulent particle flux-density appears to be due to increased turbulence activity at poloidal (toroidal) mode numbers between 200 and 400 (50 and 100) and $0.95 \leq \psi_N \leq 0.99$.
This corresponds to $k_\theta \rho_i \approx 0.3$ and is consistent with our earlier observation in Fig. \ref{fig:dphi_spectra} (f) (Sec. \ref{subsec:turbulence}).
The peaks of the particle flux in the reference simulation without RMPs that exist at $\psi_N \gtrsim 0.98$ and higher toroidal mode numbers (between 100 and 150) in Fig. \ref{fig:flux_per_mode} (a) are reduced in the simulation with RMP field in Fig. \ref{fig:flux_per_mode} (b).

\begin{figure}
 \centering
 \includegraphics[width=\textwidth]{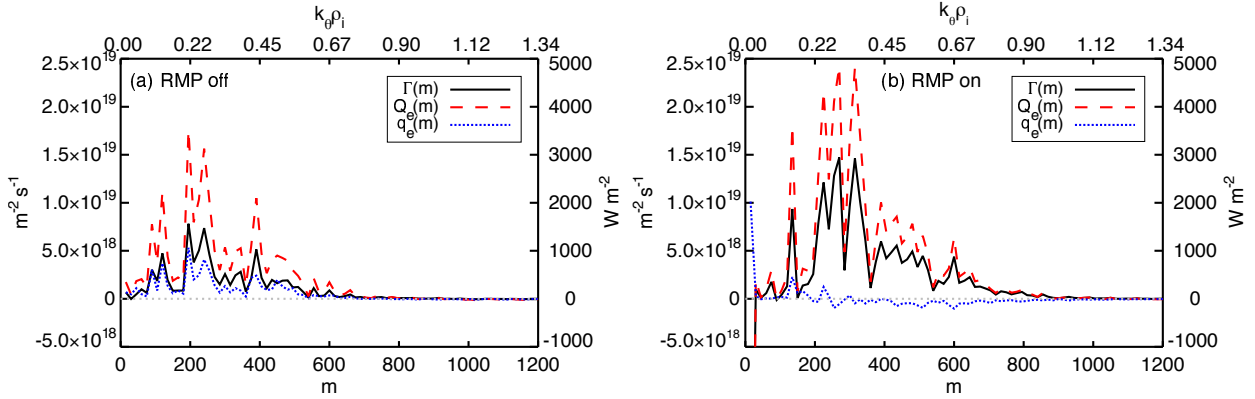}
 \caption{Turbulent particle and heat flux density as functions of the poloidal mode number $m$ at $\psi_N=0.97$: (a) without RMP field; (b) with RMP field.}
 \label{fig:flux_per_mode_1d}
\end{figure}

The turbulent electron heat flux density $q_t(\psi_N,m)$ is shown in figures \ref{fig:flux_per_mode} (c) (without RMP field) and (d) (with RMP field).
When compared to Fig. \ref{fig:flux_per_mode} (c), Fig. \ref{fig:flux_per_mode} (d) exhibits inward heat flux at $0.96 \lesssim \psi_N \lesssim 0.98$  at higher mode numbers $m\gtrsim 200$ (or $n\gtrsim 50$), resulting in suppression of the total electron heat flux.
This does not imply, though, that the total energy flux density $Q_t$ is reduced by the RMP field since it still contains the convective energy transport associated with the particle flux.
In fact, both $Q_t$ and $\Gamma_t$ are increased by the RMP field.
However, the RMP-driven enhancement of $\Gamma_t$ is relatively greater than that of $Q_t$ so that the heat flux is reduced.
This is illustrated at the flux-surface $\psi_N=0.97$ in Figs. \ref{fig:flux_per_mode_1d} (a) (without RMP field) and (b) (with RMP field).
Both the particle and total energy flux densities are increased by RMPs around $m\sim 300$ ($k \rho_i \sim 0.3$).
But the relative increase of $\Gamma_t$ is greater than $Q_t$, resulting in the reduction of the heat flux by RMPs.

\begin{figure}
 \centering
 \includegraphics[width=\textwidth]{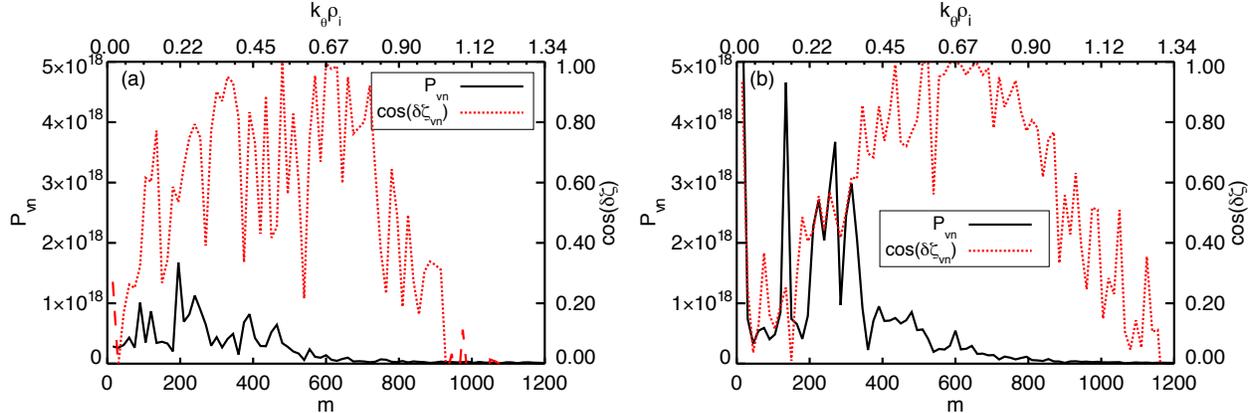}
 \caption{Cross-power $P_{vn}$ and cosine of the cross-phase $\delta\zeta_{vn}$ between radial $\exb$ velocity and density perturbations at $\psi_N=0.97$ without RMP field (a) and with RMP field (b).}
 \label{fig:cross_spectrum}
\end{figure}

In the context of the cross-spectral density, there are two routes leading to increased transport, increased cross-power or larger cosine of the cross-phase with respect to the radial $\exb$-velocity fluctuation.
Both have been discussed in the context of flow-shear suppression of turbulent transport \cite{biglari_shearflow_and_turbulence,terry_turbsuppression_zf}.
In the simulations examined for this article, the RMP-induced changes of turbulent transport are more strongly correlated with the cross-power than the cross-phase [see Fig. \ref{fig:cross_spectrum}].
Indeed, upon closer inspection of $\Gamma_t$ on the flux-surface $\psi_N=0.97$, one finds that the cosine of the cross-phase in the case with RMP field [Fig. \ref{fig:cross_spectrum} (b)] is even slightly lower than without RMPs at the mode numbers that show the largest increase in $P_{vn}$ ($m\lesssim 400$).
Only at $m \gtrsim 400$, the cosine of the cross-phase is larger with RMP field than without. 

Our analysis of the cross-spectral density of the turbulent potential, density and electron temperature fluctuations reveals that the RMP-driven increase of the turbulent particle flux is accompanied by a shift of turbulence activity to lower toroidal and poloidal mode numbers (i.e. larger scales) at $0.97 \lesssim \psi_N \lesssim 0.99$.
The suppression of the electron heat flux is primarily due to higher poloidal mode numbers ($m\gtrsim 200$).
In the electron heat-flux, no obvious predator-prey type correlation between these changes and the changes in the $\exb$ shearing rate is found.
The shift to lower turbulence mode numbers might be measurable in experiments as a shift of turbulence frequencies (provided the local $\exb$ rotation is known) or directly by imaging or scattering.

%
\section{Summary and Discussion}\label{sec:summary}

Simulations of consistent neoclassical and turbulent transport in a DIII-D H-mode plasma with resonant magnetic perturbations have been performed with the global total-f gyrokinetic PIC code XGC in order to study density-pump out and electron heat confinement in the presence of RMP fields.
The RMP field used in the XGC simulations has been computed with M3D-C1 taking into account the linear two-fluid plasma response.
Our simulations reproduce two key features of tokamak discharges with RMP field:
i) increased radial particle transport in the pedestal region that is sufficient for pumping out 50\% of the pedestal plasma in less than 100 ms;
and ii) the preservation of the electron heat transport barrier in the pedestal.
In fact, we observe complete suppression of the electron heat flux in the steepest part of the edge pedestal in the XGC simulation.
Thus, density pump-out occurs despite the presence of intact KAM surfaces in most of the pedestal ($\psi_N \lesssim 0.98$) -- similar to the experimental observation in Ref. \onlinecite{nazikian_2015} -- while electron heat remains well confined.
The suppression of the electron heat flux at $0.96<\psi < 0.98$ in our simulations suggests that the temperature pedestal may become steeper due to the RMP field, which is consistent with some experiments \cite{evans_2006}.
One caveat is, though, that our simulations are much shorter than the experimental pump-out time, i.e., they predict only the slope of the plasma profile evolution, not the steady profiles under RMP field.

The observed changes in particle and electron heat transport are correlated with increased density and decreased electron temperature fluctuation levels.
While in both simulations, with and without RMP field, ion-drift direction turbulence is active radially inside the pedestal shoulder ($\psi_N \lesssim 0.94$), and electron-drift direction turbulence is active in the steep pedestal region, the electron-directional modes are intensified by the RMP field at lower frequency, which is consistent with the enhanced particle flux.
Analyses of the cross-spectral density with the radial $\exb$ velocity and the density and temperature fluctuations show that the RMP-driven enhanced particle flux is due to an increase of the cross-power between the radial $\exb$ velocity and the density fluctuations, not from their cross-phase.
At $0.97 \lesssim \psi_N \lesssim 0.99$, a noticeable shift of the cross-power to lower poloidal and toroidal mode numbers is found, which might be measurable in experiments as a drop in the frequency of the density perturbations.

While the present work represents an important step forward in the research of RMP-driven transport and qualitatively reproduces important features, electromagnetic instabilities, a nonlinear RMP penetration model, and the self consistent evolution of the magnetic geometry are still excluded and left for future work.
Reference \onlinecite{kotschenreuther_2019} used the ``fingerprints'' technique to identify the instabilities that cause transport losses in the pedestal of many of today’s experiments, which includes micro-tearing modes (MTMs).
An implication from that work is that the electron thermal diffusion coefficient produced by MTMs could be up to $\chi_e\sim 0.3 m^2/s$, which is comparable to the maximal $\chi_e$ in the pedestal obtained from the present electrostatic simulation.
Thus, if MTM transport is active in the present edge plasma, it could have a significant impact on our conclusions.
However, the survival of MTMs in the presence of RMPs has not been observed in experiments yet and will be studied with XGC in the future.

We note here that the effective transport coefficients discovered from gyrokinetic RMP-driven transport studies, such as the present one, could be coupled with reduced model studies, such as the recent work in Ref. \citenum{hu_2019}, to allow for rapid analysis of a wide range of experimental conditions and to study experimental trends such as the density dependence of ELM suppression and pump-out.

%
\section*{Acknowledgments}
The work was performed at Princeton Plasma Physics Laboratory, which is managed by Princeton University under Contract No. DE-AC02-09CH11466, with support from the DIII-D facility under Contract No. DE-FC02-04ER54698.
Funding for this work was provided through the Scientific Discovery through Advanced Computing (SciDAC) program by the U.S. Department of Energy Office of Advanced Scientific Computing Research and the Office of Fusion Energy Sciences under Contract No. DE-AC02-09CH11466.

An award of computer time was provided by the Innovative and Novel Computational Impact on Theory and Experiment (INCITE) program. This research used resources of the Argonne Leadership Computing Facility and the National Energy Research Scientific Computing Center (NERSC), which are U.S. Department of Energy Office of Science User Facilities supported under contracts DE-AC02-06CH11357 and DE-AC02-05CH11231, respectively. 

\textbf{Disclaimer:} The publisher, by accepting the article for publication, acknowledges that the United States Government retains a non-exclusive, paid-up, irrevocable, world-wide license to publish or reproduce the published form of this manuscript, or allow others to do so, for United States Government purposes.
This report was prepared as an account of work sponsored by an agency of the United States Government.  Neither the United States Government nor any agency thereof, nor any of their employees, makes any warranty, express or implied, or assumes any legal liability or responsibility for the accuracy, completeness, or usefulness of any information, apparatus, product, or process disclosed, or represents that its use would not infringe privately owned rights. Reference herein to any specific commercial product, process, or service by trade name, trademark, manufacturer, or otherwise, does not necessarily constitute or imply its endorsement, recommendation, or favoring by the United States Government or any agency thereof. The views and opinions of authors expressed herein do not necessarily state or reflect those of the United States Government or any agency thereof.


%


\newpage
\renewcommand{\refname}{ }
\bibliographystyle{aipnum4-1}
\bibliography{references}


\end{document}